\documentclass[preprint,12pt,numbers,sort]{elsarticle}
\usepackage{graphicx}
\usepackage{epsfig}
\usepackage{epstopdf}
\usepackage{dcolumn}
\usepackage{bm}
\usepackage{amsmath}
\usepackage{color}
\usepackage{amssymb}

\usepackage{enumitem}
\usepackage{xspace}

\newcommand{\TII}{t^\mathrm{II}_0}
\newcommand{\TIII}{t^\mathrm{III}_0}
\newcommand{\tii}{$\TII$\xspace}
\newcommand{\tiii}{$\TIII$\xspace}

\newcommand{\gras}[1]{\boldsymbol{#1}}

\journal{Physics Letters B}

\begin{document}

\begin{frontmatter}

\title{Isospin-symmetry breaking in masses of $N\simeq Z$ nuclei}

\author{P. B\c{a}czyk$^a$, J. Dobaczewski$^{a,b,c,d}$, M. Konieczka$^a$,
W. Satu{\l}a$^{a,d}$, T. Nakatsukasa$^e$, K. Sato$^f$}
\address{%
$^a$Institute of Theoretical Physics, Faculty of Physics, University of Warsaw, ul. Pasteura 5, PL-02-093 Warsaw, Poland \\
$^b$Department of Physics, University of York, Heslington, York YO10 5DD, United Kingdom \\
$^c$Department of Physics, P.O. Box 35 (YFL), University of Jyv\"askyl\"a, FI-40014  Jyv\"askyl\"a, Finland \\
$^d$Helsinki Institute of Physics, P.O. Box 64, FI-00014 University of Helsinki, Finland \\
$^e$Center for Computational Sciences, University of Tsukuba, Tsukuba 305-8577, Japan \\
$^f$Department of Physics, Osaka City University, Osaka 558-8585, Japan}

\begin{abstract}
Effects of the isospin-symmetry breaking (ISB) beyond mean-field
Coulomb terms are systematically studied in nuclear masses near the
$N=Z$ line. The Coulomb exchange contributions are calculated
exactly. We use extended Skyrme energy density functionals (EDFs)
with proton-neutron-mixed densities, to which we add new terms
breaking the isospin symmetry. Two parameters associated with the new
terms are determined by fitting mirror and triplet displacement
energies (MDEs and TDEs) of isospin multiplets. The new EDFs
reproduce MDEs for the $T=\frac12$ doublets and $T=1$ triplets, and
TDEs for the $T=1$ triplets. Relative strengths of the obtained
isospin-symmetry-breaking terms {\em are not} consistent with the
differences in the $NN$ scattering lengths, $a_{nn}$, $a_{pp}$, and
$a_{np}$. Based on low-energy experimental data, it seems thus
impossible to delineate the strong-force ISB effects from
beyond-mean-field Coulomb-energy corrections.
\end{abstract}


\end{frontmatter}

\section{Introduction}
\label{Introduction}

Similarity between the neutron-neutron ($nn$), proton-proton ($pp$),
and neutron-proton ($np$) nuclear forces, commonly known as their
charge independence, has been well established experimentally already
in 1930's, leading to the concept of the isospin symmetry introduced
by Heisenberg~\cite{(Hei32b)} and Wigner~\cite{(Wig37)}. Since then,
the isospin symmetry has been tested and widely used in theoretical
modelling of atomic nuclei, with its explicit violation generated by
the Coulomb interaction. In addition, there also exists firm
experimental evidence in the nucleon-nucleon ($NN$) scattering data
that the interaction contains small isospin-symmetry-breaking (ISB)
components. The differences in the $NN$ phase shifts indicate that
the $nn$ interaction, $V_{nn}$, is about 1\% stronger than the
$pp$ interaction, $V_{pp}$, and that the $np$ interaction,
$V_{np}$, is about 2.5\% stronger than the average of $V_{nn}$ and
$V_{pp}$~\cite{(Mac01a)}. These effects are called charge-symmetry breaking
(CSB) and charge-independence breaking (CIB), respectively. In this
paper, we show that the manifestation of the CSB and CIB in nuclear
masses can systematically be accounted for by the extended nuclear
density functional theory (DFT).

The charge dependence of the nuclear strong force fundamentally
originates from mass and charge differences between $u$ and $d$
quarks. The strong and electromagnetic interactions among these
quarks give rise to the mass splitting among the baryonic and mesonic
multiplets. The neutron is slightly heavier than the proton. The
pions, which are the Goldstone bosons associated with the chiral
symmetry breaking and are the primary carriers of the nuclear force
at low energy, also have the mass splitting. The strong-force CSB
mostly originates from the difference in masses of protons and
neutrons, leading to the difference in the kinetic energies and
influencing the one- and two-boson exchange. On the other hand, the
major cause of the strong-force CIB is the pion mass splitting. For
more details, see Refs.~\cite{(Mac01a),(Mil95)}.

The Coulomb force is, of course, the major source of ISB in nuclei.
In the nuclear DFT, the Coulomb interaction is treated on the
mean-field level. Contrary to the atomic DFT, where the exchange and
correlation effects are usually treated together~\cite{(Par89)}, in
nuclei, the exchange term can be evaluated exactly, as is the case in
the present study, but the correlation terms are simply disregarded.
Therefore, the ISB terms that we introduce below are meant to describe
both the strong-force and Coulomb-correlation effects jointly.

The isospin formalism offers a convenient classification of different
components of the nuclear force by dividing them into four distinct
classes. Class-I isoscalar forces are invariant under any rotation in
the isospin space. Class-II isotensor forces break the charge
independence but are invariant under a rotation by $\pi$ with respect
to the $y-$axis in the isospace preserving therefore the charge
symmetry. Class-III isovector forces break both the charge
independence and the charge symmetry, and are symmetric under
interchange of two interacting particles. Finally, forces of class~IV
break both symmetries and are anti-symmetric under the interchange of
two particles. This classification was originally proposed by Henley
and Miller~\cite{(Mil95),(Hen79)} and subsequently used in the
framework of potential models based on boson-exchange formalism, like
CD-Bonn~\cite{(Mac01a)} or AV18~\cite{(Wir13a)}. The CSB and CIB were
also studied in terms of the chiral effective field
theory~\cite{(Wal01a),(Epe09a)}. So far, the Henley-Miller
classification has been rather rarely utilized within the nuclear
DFT~\cite{(Suz93a),(Bro00b)}, which is usually based on the
isoscalar strong forces.

The most prominent manifestation of the
ISB is in the mirror displacement energies (MDEs) defined as the
differences between binding energies of mirror nuclei:
\begin{equation}
\mathrm{MDE}=BE(T,T_z\mbox{=}-T)-BE(T,T_z\mbox{=}+T). \label{eq:MDE}
\end{equation}
A systematic study by Nolen and Schiffer~\cite{(Nol69)} showed that
the MDEs cannot be reproduced by using models involving mean-field Coulomb
interaction as the only source of the ISB, see also
Refs.~\cite{(Orm89a),(Bro00b),(Kan13),(Col98)}. Another source of
information on the ISB are the so-called triplet displacement
energies (TDEs):
\begin{equation}
\mathrm{TDE}=BE(T\mbox{=}1,T_z\mbox{=}-1)+BE(T\mbox{=}1,T_z\mbox{=}+1)
-2BE(T\mbox{=}1,T_z\mbox{=}0),\label{eq:TDE}
\end{equation}
which are measures of the binding-energy curvatures within the
isospin trip\-lets. The TDEs cannot be reproduced by means of
conventional approaches disregarding nuclear CIB forces either,
see~\cite{(Orm89a),(Orm17)}. In the above definitions of MDEs
and TDEs, the binding energies are negative ($BE<0$) and the proton
(neutron) has isospin projection of $t_z=-\frac12(+\frac12)$,
that is, $T_z=\frac12(N-Z)$.

\begin{figure}[ht!]
\centerline{\includegraphics[width=0.8\columnwidth]{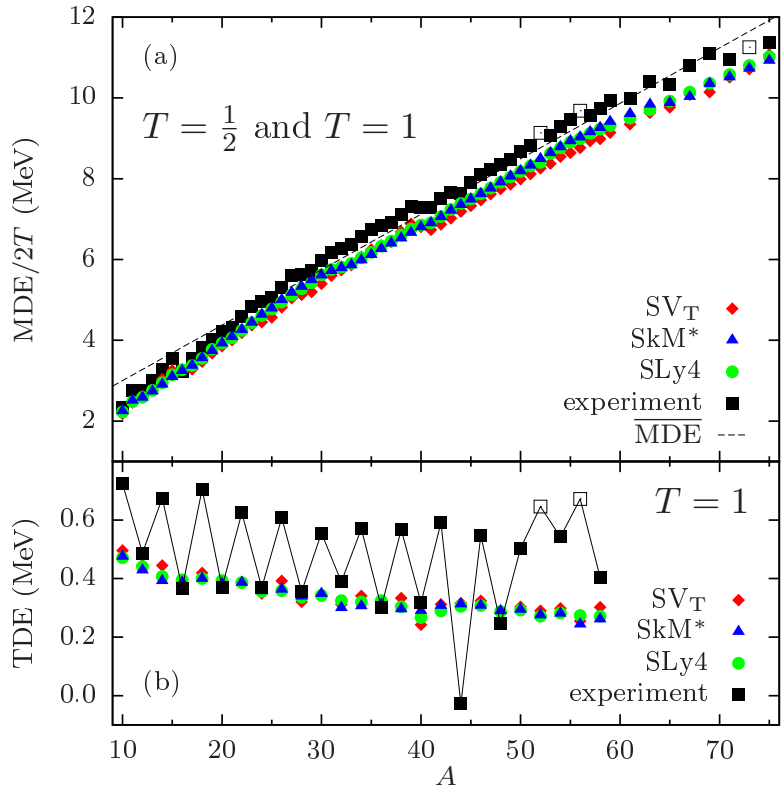}}
\caption{(Color online) Calculated (no ISB terms) and experimental
values of MDEs (a) and TDEs (b). The values of MDEs for triplets are
divided by two to fit in the plot. Thin dashed line shows the average
linear trend of experimental MDEs in doublets, defined as
$\overline{MDE} = 0.137A + 1.63$ (in MeV). Measured values of binding
energies were taken from Ref.~\cite{(Wan12a)} and the excitation
energies of the $T=1$, $T_z=0$ states from Ref.~\cite{(ensdf_url)}.
Open squares denote data that depend on masses derived from
systematics~\cite{(Wan12a)}.
}
\label{fig:MDE_TDE}
\end{figure}

In Fig.~\ref{fig:MDE_TDE}, we show MDEs and TDEs calculated fully
self-consistently using three different standard Skyrme
energy-density functionals (EDFs): SV$_{\rm
T}$~\cite{(Bei75b),(Sat14g)}, SkM$^*$~\cite{(Bar82b)}, and
SLy4~\cite{(Cha98a)}. Details of the calculations, performed using
code HFODD~\cite{(Sch17)}, are presented in the Supplemental
Material~\cite{suppl-ted}\nocite{(Bac17a),(Sat12b)}. In
Fig.~\ref{fig:MDE_TDE}(a), we clearly see that the values of obtained
MDEs are systematically lower than the experimental ones by about
10\%. Even more spectacular discrepancy appears in
Fig.~\ref{fig:MDE_TDE}(b) for TDEs, namely, for the $A=4n$ triplets
their values are nicely reproduced by the mean-field Coulomb effects,
however, the characteristic staggering pattern seen in experiment is
entirely absent. (See below for the discussion regarding the outlier
case of $^{44}$V.) It is also very clear that the calculated MDEs and
TDEs, which are specific differences of binding energies, are
independent of the choice of Skyrme EDF parametrization, that is, of
the isoscalar part of the EDF.

\section{Methods}
\label{Methods}

We aim at a comprehensive study of MDEs and TDEs based on the extended
Skyrme $pn$-mixed DFT~\cite{(Sat13d),(She14a),(Sch17)},
which includes zero-range class-II and III forces.
We consider the following zero-range
interactions of class~II and III with two new low-energy
coupling constants \tii and \tiii~\cite{(Bac16a)}:
\begin{eqnarray}
\hat{V}^{\rm{II}}(i,j) & = &
t_0^{\rm{II}}\, \delta\left(\gras{r}_i - \gras{r}_j\right)
\left[3\hat{\tau}_3(i)\hat{\tau}_3(j)-\hat{\vec{\tau}}(i)\circ\hat{\vec{\tau}}(j)\right],
\label{eq:Skyrme_classII}\\
\hat{V}^{\rm{III}}(i,j) & = &
t_0^{\rm{III}}\, \delta\left(\gras{r}_i - \gras{r}_j\right)
\left[\hat{\tau}_3(i)+\hat{\tau}_3(j)\right] .
\label{eq:Skyrme_classIII}
\end{eqnarray}
The corresponding contributions to EDF read:
\begin{eqnarray}
\mathcal{H}_{\rm{II}} &=& \frac{1}{2}t_0^{\rm{II}}
(\rho_n^2+\rho_p^2-2\rho_n\rho_p-2\rho_{np}\rho_{pn}\notag\\
&&-\gras{s}_{n}^2-\gras{s}_{p}^2+2\gras{s}_{n}\cdot\gras{s}_{p}+2\gras{s}_{np}\cdot\gras{s}_{pn}),\label{eq:ED_classII}
\\
\mathcal{H}_{\rm{III}} &=& \frac{1}{2}t_0^{\rm{III}}
\left(\rho_n^2-\rho_p^2 - \gras{s}_{n}^2+\gras{s}_{p}^2\right),
\label{eq:ED_classIII}
\end{eqnarray}
where $\rho$ and $\gras{s}$ are scalar and spin (vector) densities,
respectively. Inclusion of the spin exchange terms in
Eqs.~(\ref{eq:Skyrme_classII}) and (\ref{eq:Skyrme_classIII}) leads
to a trivial rescaling of coupling constants \tii and \tiii,
see~\cite{(Bac16a)}. Hence, these terms were disregarded.

Contributions of class-III force to EDF (\ref{eq:ED_classIII}) depend
on the standard $nn$ and $pp$ densities and, therefore, can be taken
into account within the conventional $pn$-separable DFT
approach~\cite{(Suz93a),(Bro00b)}. In contrast, contributions of
class-II force (\ref{eq:ED_classII}) depend explicitly on the mixed
densities, $\rho_{np}$ and $\gras{s}_{np}$, and require the use of
$pn$-mixed DFT~\cite{(Sat13d),(She14a)}.

We implemented the new terms of the EDF in the code
HFODD~\cite{(Sch17)}, where the isospin degree of freedom is
controlled within the isocranking
method~\cite{(Sat01a),(Sat01b),(Sat13d)} -- an analogue of the
cranking technique that is widely used in high-spin physics. The
isocranking method allows us to calculate the entire isospin
multiplet, $T$, by starting from an isospin-aligned state
$|T,T_z\mbox{=}T\rangle$ and isocranking it around axes tilted with
respect to the $z$-axis. In particular, the isocranking around the
$x$-axis (or eqivalently around the $y$-axis) allows us to reach the
state with $\langle{\hat T}_z\rangle\simeq0$.

A rigorous treatment of the isospin symmetry within the $pn$-mixed
DFT formalism requires full, three-dimensional isospin projection,
which is currently under development. However, for the purpose of
calculating values of TDEs, we can perform the isospin
projection in the following way. First, we treat the standard effects
of the isospin mixing caused by the Coulomb interaction at the
mean-field level, cf.~\cite{(Sat10)}. That is, we consider
mean-field states of nuclei with $(T,T_z\mbox{=}{\pm}1)$ as having
approximately good isospin. Then, the only states that need special
attention are those with $(T,T_z\mbox{=}0)$, which are obtained by the
isocranking technique.

Let us denote by
$|\Psi_{T,Tz}\rangle$= $\{|\Psi_{1,-1}\rangle,|\Psi_{1,0}\rangle,|\Psi_{1,1}\rangle\}$
the wavefunctions of the triplet of states. For the $A=4n+2$ triplets,
these wave functions can be very simply written as
\begin{eqnarray}
|\Psi_{1,-1} \rangle&=&a^+_{p\uparrow}a^+_{p\downarrow}|0\rangle,  \nonumber \\
|\Psi_{1,0}~~\rangle&=&\tfrac{1}{\sqrt{2}}\left(a^+_{n\uparrow}a^+_{p\downarrow}+a^+_{p\uparrow}a^+_{n\downarrow}\right)|0\rangle, \label{eq:Psi} \\
|\Psi_{1,1}~~\rangle&=&a^+_{n\uparrow}a^+_{n\downarrow}|0\rangle,  \nonumber
\end{eqnarray}
where $\uparrow$ and $\downarrow$ denote pairs of Kramers-degenerate deformed states,
and $|0\rangle$ denotes the $T=0$ core of $A=4n$ particles.

Similarly, the $x$-isocranked state reads
\begin{eqnarray}
|\Psi_{++}\rangle&=&a^+_{+\uparrow}a^+_{+\downarrow}|0\rangle =
\tfrac{1}{2}|\Psi_{1,-1}\rangle+\tfrac{1}{\sqrt{2}}|\Psi_{1,0}\rangle+\tfrac{1}{2}|\Psi_{1,1}\rangle,  \label{eq:Psi2}
\end{eqnarray}
where
$|+\uparrow\rangle=\tfrac{1}{\sqrt{2}}\left(|n\uparrow\rangle+|p\uparrow\rangle\right)$
and
$|+\downarrow\rangle=\tfrac{1}{\sqrt{2}}\left(|n\downarrow\rangle+|p\downarrow\rangle\right)$
are single-particle eigenstates of the Pauli matrix $\tau_x$. Since
all terms in the Hamiltonian are diagonal in $T_z$, we then have the
binding energy of the isocranked state as
$BE_{++}=\tfrac{1}{4}BE(T$$=$$1,T_z$$=$$-1)+\tfrac{1}{2}BE(T$$=$$1,T_z$$=$$0)+\tfrac{1}{4}BE(T$$=$$1,T_z$$=$$+1)$.
When this result is inserted into Eq.~(\ref{eq:TDE}), we finally
obtain\footnote{In Refs.~\cite{(Bac17a),(Bac16a),(Sat14f),(Bac17c)}, we have
erroneously used Eq.~(\ref{eq:TDE}) with $BE(T\mbox{=}1,T_z\mbox{=}0)$ replaced
by $BE_{++}$, which resulted in numerical values of
TDEs being twice too small, cf.~Eq.~(\ref{eq:TDE2}), and in incorrect
values of the adjusted coupling constants \tii.}
\begin{gather}
\mathrm{TDE}=2\Big[BE(T\mbox{=}1,T_z\mbox{=}-1)+BE(T\mbox{=}1,T_z\mbox{=}+1)
-2BE_{++}\Big].\label{eq:TDE2}
\end{gather}
For the $A=4n$ triplets, the derivation is slightly more involved,
but the same result (\ref{eq:TDE2}) holds. In this way, TDEs of the
isospin-projected triplets can be determined from energies of three
Slater determinants: $|\Psi_{1,-1}\rangle$, $|\Psi_{++}\rangle$, and
$|\Psi_{1,1}\rangle$. The procedure proposed in
Eqs.~(\ref{eq:Psi})--(\ref{eq:TDE2}) is equivalent to an exact
projection on the $N=Z$ $T=1$ component of the isocranked
wavefunction, which amounts to removing its dispersion in $T_z$.

Physically relevant values of \tii and \tiii turn out to be fairly
small~\cite{(Bac16a)}, and thus the new terms do not impair the
overall agreement of self-consistent results with the standard
experimental data. Moreover, calculated MDEs and TDEs depend on \tii
and \tiii almost linearly, and, in addition, MDEs (TDEs) depend very
weakly on \tii (\tiii) \cite{(Bac16a),suppl-ted}. This allows us to use the
standard linear regression method, see,
e.g.~Refs.~\cite{(Toi08b),(Dob14b)}, to independently adjust \tii and
\tiii to experimental values of TDEs and MDEs, respectively. See
Supplemental Material~\cite{suppl-ted} for detailed description of
the procedure. Coupling constants \tii and \tiii resulting from such
an adjustment are collected in Table~\ref{tab:t-param}. We have
performed adjustements to masses tabulated in AME2012~\cite{(Wan12a)};
in this way, below we can test our predictions by comparing the
results to those tabulated in AME2016~\cite{(Wan17)}.

\begin{table}[ht!]
\centering
\caption{Coupling constants \tii and \tiii and their uncertainties
obtained in this work for the Skyrme EDFs SV$_{\rm T}^{\text{ISB}}$,
SkM*$^{\text{ISB}}$, and SLy4$^{\text{ISB}}$. In the last row we show their
corresponding ratios.}
\label{tab:t-param}
\begin{tabular}{r@{~~~}l@{~~~}l@{~~~}l}\hline  \rule{0mm}{3.5mm}
                   & ~~~SV$_{\rm T}^{\text{ISB}}$  & ~~~SkM*$^{\text{ISB}}$& ~~~SLy4$^{\text{ISB}}$             \\ \hline  \rule{0mm}{4mm}
\tii  (MeV fm$^3$) & $ \phantom{-}4.6\pm1.6$ &   $ \phantom{-5.}7   \pm 4   $ & $ \phantom{-5.}6   \pm 4   $ \\
\tiii (MeV fm$^3$) & $           -7.4\pm1.9$ &   $             -5.6 \pm 1.4 $ & $             -5.6 \pm 1.1 $ \\
\hline  \rule{0mm}{3.5mm}
\tii/\tiii         & $           -0.6\pm0.3$ &   $             -1.3 \pm 0.8 $ & $             -1.1 \pm 0.7 $ \\ \hline
\end{tabular}
\end{table}

\section{Results}
\label{Results}

\begin{figure}[ht!]
\centerline{\includegraphics[width=0.8\columnwidth]{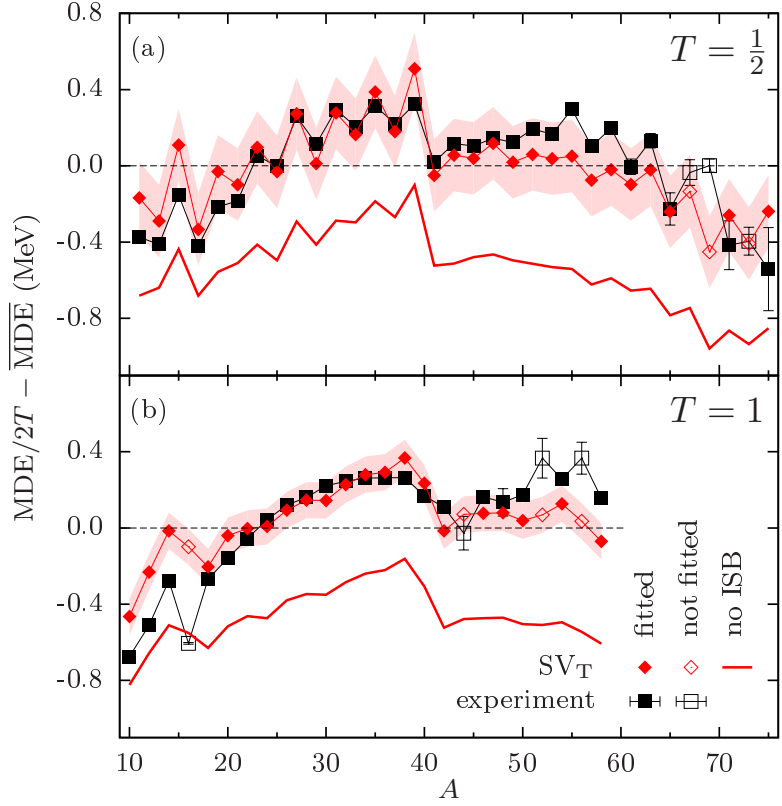}}
\caption{(Color online)
Calculated and experimental~\cite{(Wan12a)} values of MDEs for the $T=\frac12$ (a)
and $T=1$ (b) mirror nuclei, shown with respect to the average linear
trend defined in Fig.~\protect\ref{fig:MDE_TDE}. Calculations were
performed for functional SV$_{\rm T}^{\text{ISB}}$.
Shaded bands show theoretical uncertainties. Experimental error bars
are shown only when they are larger than the corresponding symbols.
Full (open) symbols denote data points included in (excluded from)
the fitting procedure.
}
\label{fig:MDE_SV}
\end{figure}

\begin{figure}[ht!]
\centerline{\includegraphics[width=0.8\columnwidth]{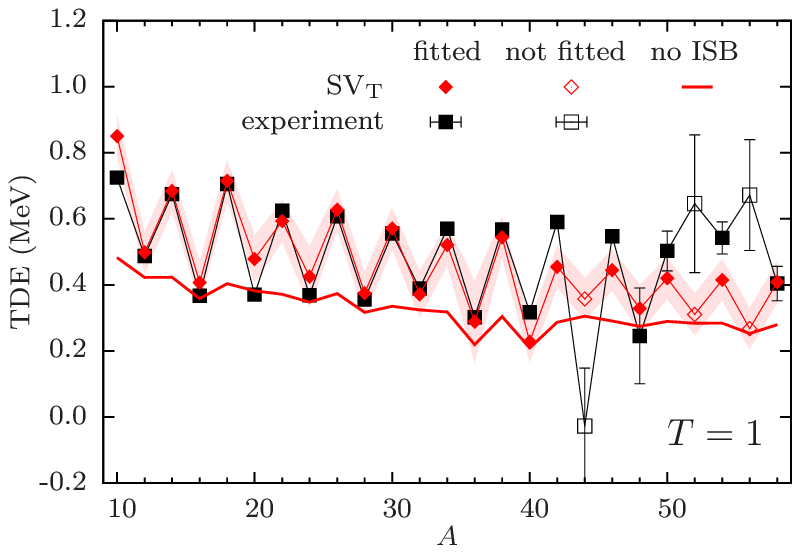}}
\caption{(Color online) Same as in Fig.~\protect\ref{fig:MDE_SV}
but for the $T=1$ TDEs with no linear trend subtracted.}
\label{fig:TDE_SV}
\end{figure}

In Fig.~\ref{fig:MDE_SV}, we show values of MDEs calculated within
our extended DFT formalism for the Skyrme SV$_{\rm T}^{\text{ISB}}$
EDF. By subtracting an overall linear trend (as defined in
Fig.~\ref{fig:MDE_TDE}) we are able to show results in an extended
scale, for which a detailed comparison with experimental data is
possible. In Fig.~\ref{fig:TDE_SV}, we show the corresponding
SV$_{\rm T}^{\text{ISB}}$ values of TDEs, whereas complementary
results obtained for the Skyrme SkM*$^{\text{ISB}}$ and
SLy4$^{\text{ISB}}$ EDFs are collected in the Supplemental
Material~\cite{suppl-ted}. Here, we concentrate on the results given
by the Skyrme SV$_{\rm T}^{\text{ISB}}$ EDF, because it is the only
one based on averaging a two-body pseudopotential (without
density-dependent terms), and it is thus free from unwanted
self-interaction contributions~\cite{(Tar14a)}.

It is gratifying to see that the calculated values of MDEs closely
follow the experimental $A$-dependence, see Fig.~\ref{fig:MDE_SV}. It
is worth noting that a single coupling constant \tiii reproduces both
$T=\frac12$ and $T=1$ MDEs, which confirms conclusions of
Refs.~\cite{(Suz93a),(Bro00b)}. In addition, for the $T=\frac12$
MDEs, Fig.~\ref{fig:MDE_SV}(a), the SV$_{\rm T}^{\text{ISB}}$ results
nicely reproduce (i) changes in experimental trend that occur at
$A=15$ and 39, (ii) staggering pattern between $A=15$ and 39, and
(iii) quenching of staggering between $A=41$ and 53 (the $f_{7/2}$
nuclei).

\begin{figure}[ht!]
\centerline{\includegraphics[width=0.8\columnwidth]{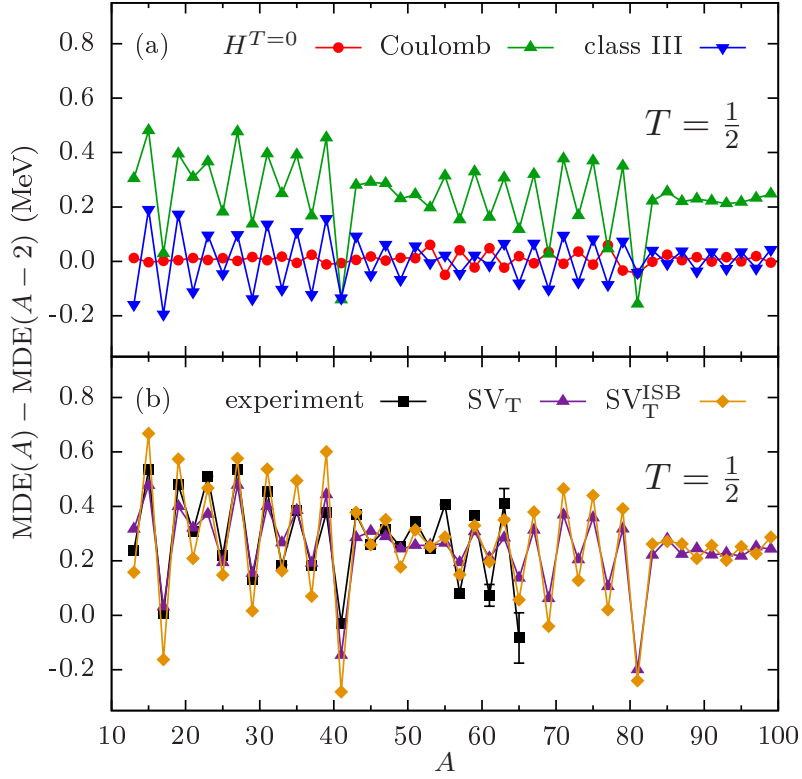}}
\caption{(Color online)
Staggering of the calculated $T=\frac12$ MDEs: MDE($A)$$-$MDE($A$$-$$2$).
(a) Separate contributions of the isoscalar ($H^{\text{T=0}}$),
Coulomb, and class-III terms, determined for functional SV$_{\rm
T}^{\text{ISB}}$. (b) Results obtained for functionals SV$_{\rm T}$
(no ISB) and SV$_{\rm T}^{\text{ISB}}$ (fitted) compared with
experimental data~\cite{(Wan12a)}.
}
\label{fig:DMDE_SV}
\end{figure}

We note that these features are already present in the SV$_{\rm T}$
results without the ISB terms, that is, for the mean-field Coulomb
interaction. On top of the Coulomb force, the class-III force is
essential in bringing the magnitude of MDEs up to the experimental
values, but also in simultaneously {\it increasing} the staggering
pattern given by the Coulomb interaction. This is illustrated in
Fig.~\ref{fig:DMDE_SV}(a), where we plotted differences
MDE($A)$$-$MDE($A$$-$2), separately for the contributions coming from
the isoscalar, Coulomb, and class-III terms of the functional.

On the one hand, we see that the Coulomb force alone always induces
staggering of MDEs, except in the $f_{7/2}$ and $g_{9/2}$ nuclei,
where this part of the staggering disappears almost completely. On the
other hand, the class-III force induces a (smaller) {\em in-phase}
staggering in all systems. Because of the self-consistency, the
isoscalar terms also show some small out-of-phase staggering, which
is a result of strong cancellation between fairly large
kinetic-energy and Skyrme-force contributions. In
Fig.~\ref{fig:DMDE_SV}(b), we showed differences
MDE($A)$$-$MDE($A$$-$2) calculated with and without the ISB terms,
compared with experimental values~\cite{(Wan12a)}. This figure also
shows results of our calculations extrapolated up to $A=99$.

For all three functionals our results correctly describe the
$A$-dependence, and lack of staggering, of the $T=1$ MDEs, see
Fig.~\ref{fig:MDE_SV}(b) and~\cite{suppl-ted}. Coming to the
discussion of TDEs, it is even more gratifying to see in
Fig.~\ref{fig:TDE_SV} that our $pn$-mixed calculations, with one
adjusted class-II coupling constant, \tii, describe absolute values
and staggering of TDEs quite well. By including the class-II force into the SV$_{\rm T}$ parametrization,
the overall rms deviations of TDEs decreases from 190 to 69\,keV.
In fact, the improvement comes from the decrease of the rms deviations for the $A=4n+2$
triplets, from 250 to 75\,keV, wheres those for the $A=4n$ triplets change very little, from 58 to 60\,keV.

We also note that (i) in our approach, the staggering of MDEs in the
$f_{7/2}$ shell is increased by the ISB terms, leading to an
agreement with data, Fig.~\ref{fig:DMDE_SV}(b), whereas in the
results of Ref.~\cite{(Kan13)} this staggering is decreased by the
ISB forces. Also, (ii) in light of our results, the standard
interpretation of a phenomenological term in Coulomb
energies~\cite{(Nol69),(Shl78a)} that is proportional to $(-1)^Z$, is
not correct. This term was introduced to remove effects of the proton
Coulomb self-energies. Our calculations treat the Coulomb-exchange
terms exactly, and thus are free from self-energies; nevertheless,
these exact Coulomb energies do not show any staggering of the TDEs,
Fig.~\ref{fig:TDE_SV}. Therefore, our results disprove the motivation
for introducing a phenomenological $(-1)^Z$ term that generates a
uniform staggering of MDEs and TDEs. And (iii) for the
SkM*$^{\text{ISB}}$ and SLy4$^{\text{ISB}}$ functionals, the
staggering of the $T=\frac12$ MDEs and TDEs is less pronounced than
for SV$_{\rm T}^{\text{ISB}}$~\cite{suppl-ted} and the results
obtained without the ISB terms show almost no staggering in
$T=\frac12$ MDEs. This may suggest that the staggering of Coulomb
energies in $T=\frac12$ MDEs may be washed out by the
self-interaction contributions, which are present in the
SkM*$^{\text{ISB}}$ and SLy4$^{\text{ISB}}$ functionals.

Good agreement obtained for the MDEs and TDEs shows that the role and
magnitude of the simplest DFT ISB terms are now firmly established.
Nevertheless, some conspicuous deviations of our theoretical
predictions with respect to the experiment still remain. This
includes (i) overestimated (underestimated) values of MDEs
in lighter (heavier) multiplets, (ii) overestimated (underestimated)
staggering of MDEs in lighter (heavier) doublets, and (iii)
underestimated staggering of TDEs in
heavier triplets. This suggests that higher-order DFT ISB terms,
that is, gradient terms that would generate dependence on surface
properties, may still play some role.

It is very instructive to look at ten outliers that were excluded
from the fitting procedure. In Figs.~\ref{fig:MDE_SV}
and~\ref{fig:TDE_SV}, they are shown by open symbols. They can be
classified as (i) five outliers that depend on masses of $^{52}$Co,
$^{56}$Cu, and $^{73}$Rb, which clearly deviate from the calculated
trends for MDEs and TDEs. These masses were not directly measured but
were derived from systematics~\cite{(Wan12a)}. And (ii) two outliers that
depend on the mass of $^{44}$V, whose ground-state measurement may be
contaminated by an unresolved isomer~\cite{(Fuj13),(Mac14),(Tu16)}.
As well as (iii), large differences between experimental and
calculated values are found in MDE for $A=16$, 67 and 69. Inclusion
of these data in the fitting procedure would significantly increase
the uncertainty of adjusted coupling constants. The former two
classes of outliers, (i) and (ii), call for improving experimental
values, whereas the last one (iii) may be a result of structural
effects not included in our model.

Our results can be confronted with the state-of-the-art global
analysis performed within the shell-model~\cite{(Orm89a)}. There, the
MDEs and TDEs were very accurately described by fitting the Coulomb,
class-II, and class-III shell-model interactions, separately in five
different valence spaces between $A=10$ and 55. In addition, values
of 14 single-particle energies were also adjusted. As a result of
such a 29-parameter fit, the rms deviations between measured and
calculated values of MDEs and TDEs were obtained as 44 and 23\,keV,
respectively. This can be contrasted with our two-parameter fit in
SV$_{\rm T}^{\text{ISB}}$, resulting in the corresponding rms
deviations of 220 and 66\,keV. (Here, we have used the same set of
nuclei as that analyzed in Ref.~\cite{(Orm89a)}.) Undoubtedly, this
shows that by adding higher-order DFT ISB terms, our results can
still be improved. Nevertheless, we can conclude that both the DFT
and shell-model analyses consistently point to the necessity of using
specific ISB terms to account for masses of $N\simeq Z$ nuclei.

Having at hand a model with the ISB interactions included,
we can calculate MDEs and TDEs for more massive multiplets, and
make predictions of binding energies for neutron-deficient ($T_z =
-T$) nuclei. In particular, in Table~\ref{tab:mass_excess} we present
predictions of mass excesses of $^{52}$Co, $^{56}$Cu, and $^{73}$Rb,
whose masses were in AME2012~\cite{(Wan12a)} derived from systematics,
and $^{44}$V, whose ground-state mass measurement is uncertain.
Recently, the mass excess of $^{52}$Co was measured as
$-$34361(8)~\cite{(Xu16)} or $-$34331.6(66)~keV~\cite{(Nes17b)}.
These values are in excellent agreement with our prediction, even
though the difference between them is still far beyond the estimated
(much smaller) experimental uncertainties. We also note that a similar
excellent agreement is obtained between our mass excess of
$^{56}$Cu and that of AME2016~\cite{(Wan17)}. On the other hand,
estimates given in Ref.~\cite{(Tu16)} are outside our error bars.

\begin{table}
\centering
\caption{Mass excesses of $^{52}$Co, $^{56}$Cu, $^{73}$Rb, and $^{44}$V
obtained in this work and compared with those of AME2012~\protect\cite{(Wan12a)}
and AME2016~\protect\cite{(Wan17)}.
Our predictions were calculated as weighted averages of values obtained
from MDEs and TDEs for all three used Skyrme parametrizations.
The AME values derived from systematics are labelled with symbol \#.}
\label{tab:mass_excess}
\begin{tabular}{c@{~~~}c@{~~~~}l@{~~~~}l}\hline  \rule{0mm}{3.5mm}
& \multicolumn{3}{c}{Mass excess (keV)} \\
Nucleus & This work & AME2012~\protect\cite{(Wan12a)} & AME2016~\protect\cite{(Wan17)}\\ \hline \rule{0mm}{4mm}
$^{52}$Co & $-$34370(40) & $-$33990(200)\# & $-$34361.0(84)  \\
$^{56}$Cu & $-$38650(40) & $-$38240(200)\# & $-$38643(15)    \\
$^{73}$Rb & $-$46100(80) & $-$46080(100)\# & $-$46080(200)\# \\
$^{44}$V  & $-$23710(40) & $-$24120(180)   & $-$24120(180)   \\\hline
\end{tabular}
\end{table}

Assuming that the extracted CSB and CIB effects are, predominantly,
due to the ISB in the $^1S_0$ $NN$ scattering channel, one can
attempt relating ratio  \tii / \tiii to the experimental scattering
lengths. The reasoning follows the work of Suzuki~\textit{et
al.}~\cite{(Suz93a)}, who assumed a proportionality between the
strengths of CSB and CIB forces and the corresponding scattering
lengths~\cite{(Mil90b)}, that is, $V_{CSB} \propto \Delta
a_{CSB}=a_{pp}-a_{nn}$ and $V_{CIB} \propto \Delta
a_{CIB}=\frac12(a_{pp}+a_{nn})-a_{np}$, which, in our case, is
equivalent to $\TIII\propto-\frac12 \Delta a_{CSB}$ and $\TII\propto
\frac13 \Delta a_{CIB}$. Assuming further that the proportionality
constant are the same, and taking for the experimental values $\Delta
a_{CSB}=1.5\pm0.3\mathrm{~fm}$ and $\Delta
a_{CIB}=5.7\pm0.3\mathrm{~fm}$~\cite{(Mil90b)}, one gets the
estimate:
\begin{equation}\label{eq:ratio}
\frac{t_0^{\rm{II}}}{t_0^{\rm{III}}}=-\frac23\frac{\Delta a_{CIB}}{\Delta a_{CSB}} = -2.5 \pm 0.5.
\end{equation}
From the values of coupling constants \tii and \tiii obtain in this
work, we can obtain their ratios as given in Table~\ref{tab:t-param}.
As we can see, the ratios determined by our analysis of masses of
$N\simeq Z$ nuclei with $10\leq A\leq 75$ have a correct sign but are
2--4 times smaller than the estimate (\ref{eq:ratio}) based on
properties of the $NN$ forces deduced from the $NN$ scattering
experiments.

\section{Conclusions}
\label{Conclusions}

In this Letter, we showed that the nuclear DFT with added two new
terms related to the ISB interactions of class II and III is able to
systematically reproduce observed MDEs and TDEs of $T=\frac12$ and
$T=1$ multiplets. Adjusting only two coupling constants, \tii and
\tiii, we reproduced not only the magnitudes of the MDE and TDE but
also their characteristic staggering patterns. The obtained values of
\tii and \tiii turn out to {\em not agree} with the $NN$ ISB
interactions ($NN$ scattering lengths) in the $^1S_0$ channel. We
predicted mass excesses of $^{52}$Co, $^{56}$Cu, $^{73}$Rb, and
$^{44}$V, and for $^{52}$Co and $^{56}$Cu we obtained excellent agreement with the
recently measured values~\cite{(Xu16),(Nes17b),(Wan17)}. To better pin down
the ISB effects, accurate mass measurements of the other two nuclei
are very much called for.

Our work constitutes the first global DFT study of TDEs in the $T=1$
isomultiplets. It is based on the introduction of a single class-II
ISB term within the $pn$-mixed DFT~\cite{(Sat13d),(She14a)}. In addition, we
confirmed results of Refs.~\cite{(Suz93a),(Bro00b)}, which described
the values of MDEs by using a single class-III ISB term. We also
showed that the characteristic staggering patterns of MDEs are mostly
related to the standard Coulomb effects, whereas those of TDEs
require introducing the class-II ISB terms. Altogether, we showed
that very simple general DFT ISB terms properly account for all
currently available experimental data for MDEs and TDEs.

Finally, we note that our adjusted ISB terms probably jointly include
effects of the Coulomb correlations beyond mean field and ISB strong
interactions, along with possible many-body ISB correlations induced
by them. It does not seem reasonable to expect that low-energy
nuclear experimental data would allow for disentangling these
distinct sources of the ISB in finite nuclei. In this respect, {\it
ab initio} derivations, such as recently performed for four triplets
in Ref.~\cite{(Orm17)}, would be very important, provided they cover
the entire set of data available experimentally.

Interesting comments by Witek Nazarewicz are gratefully acknowledged.
This work was supported in part by the Polish National Science Center
under Contract Nos.\ 2014/15/N/ST2/03454 and 2015/17/N/ST2/04025, by
the Academy of Finland and University of Jyv\"askyl\"a within the
FIDIPRO program, by Interdisciplinary Computational Science Program
in CCS, University of Tsukuba, and by ImPACT Program of Council for
Science, Technology and Innovation (Cabinet Office, Government of
Japan). We acknowledge the CI\'S \'Swierk Computing Center, Poland,
and the CSC-IT Center for Science Ltd., Finland, for the allocation
of computational resources.

\bibliography{TEDMED,jacwit33}
\bibliographystyle{elsarticle-num}

\end{document}


\begin{frontmatter}

\title{Supplemental Material for: Isospin-symmetry breaking
in masses of $N\simeq Z$ nuclei}

\author{P. B\c{a}czyk$^a$, J. Dobaczewski$^{a,b,c,d}$, M. Konieczka$^a$,
W. Satu{\l}a$^{a,d}$, T. Nakatsukasa$^e$, K. Sato$^f$}
\address{%
$^a$Institute of Theoretical Physics, Faculty of Physics, University of Warsaw, ul. Pasteura 5, PL-02-093 Warsaw, Poland \\
$^b$Department of Physics, University of York, Heslington, York YO10 5DD, United Kingdom \\
$^c$Department of Physics, P.O. Box 35 (YFL), University of Jyv\"askyl\"a, FI-40014  Jyv\"askyl\"a, Finland \\
$^d$Helsinki Institute of Physics, P.O. Box 64, FI-00014 University of Helsinki, Finland \\
$^e$Center for Computational Sciences, University of Tsukuba, Tsukuba 305-8577, Japan \\
$^f$Department of Physics, Osaka City University, Osaka 558-8585, Japan}

\begin{abstract}
Effects of the isospin-symmetry breaking (ISB) beyond mean-field
Coulomb terms are systematically studied in nuclear masses near the
$N=Z$ line. The Coulomb exchange contributions are calculated
exactly. We use extended Skyrme energy density functionals (EDFs)
with proton-neutron-mixed densities, to which we add new terms
breaking the isospin symmetry. Two parameters associated with the new
terms are determined by fitting mirror and triplet displacement
energies (MDEs and TDEs) of isospin multiplets. The new EDFs
reproduce MDEs for the $T=\frac12$ doublets and $T=1$ triplets, and
TDEs for the $T=1$ triplets. Relative strengths of the obtained
isospin-symmetry-breaking terms {\em are not} consistent with the
differences in the $NN$ scattering lengths, $a_{nn}$, $a_{pp}$, and
$a_{np}$. Based on low-energy experimental data, it seems thus
impossible to delineate the strong-force ISB effects from
beyond-mean-field Coulomb-energy corrections.
\end{abstract}

\begin{keyword}
nuclear density functional theory (DFT) \sep
energy density functional (EDF) \sep
proton-neutron mixing \sep
isospin symmetry breaking (ISB) \sep
mirror displacement energy (MDE) \sep
triple displacement energy (TDE)



\end{keyword}

\end{frontmatter}

This Supplemental Material explains technical aspects of the method
presented in the Letter and provides numerical results that
complement those of the Letter. We start by providing the
reader with details concerning the choice of the basis size.  Next,
we discuss the subtleties related to the fitting procedure of the two
new coupling constants.  Finally, we present results obtained for the
MDEs and TDEs calculated with the isospin-symmetry breaking
(ISB) terms included in the SkM*$^{\text{ISB}}$~\cite{(Bar82b)} and
SLy4$^{\text{ISB}}$~\cite{(Cha98a)} Skyrme functionals that mirror
those for SV$_{\rm T}^{\text{ISB}}$~\cite{(Bei75b),(Sat14g)}, which
were presented in the Letter.

\section{Choice of the basis size}

The code HFODD~\cite{(Sch17)} used in this work solves the
Hartree-Fock (HF) equations in the Cartesian harmonic oscillator (HO)
basis. In this work, we use the spherical HO basis whose size is
determined by the number of shells, $N_0$, taken into account.
Numerical stability with respect to $N_0$ was tested in detail in
Ref.~\cite{(Bac17a)}. It turned out that the optimum choice of
number of HO shells (in terms of the trade-off between precision and
computing time) is $N_0=10$ for light nuclei ($10 \leq A \leq 30$),
$N_0=12$ for medium-mass nuclei ($31 \leq A \leq 56$) and $N_0=14$
for heavier nuclei ($A\geq 57$). In the Letter, we followed these
guidelines with exception of the $A=58$ triplet, which was
calculated with $N_0=12$ to maintain consistency with other triplets.

\section{Adjustment of the model coupling constants}

Results of fit of the model coupling constants \tii and \tiii, together
with essential elements of the fitting procedure, are given in the
Letter. In the following subsections we provide further details
concerning the methodology used. In particular, we discuss
arguments justifying the splitting of the two-dimensional (2D) fit
into two one-dimensional (1D) fits. We also discuss and
quantify sources of potential errors and uncertainties.

\subsection{Selecting HF solutions used for fitting}

Calculations of MDEs and TDEs involve ground-state
energies in even-even, odd-$A$,  and odd-odd nuclei. In the even-even
$A=4n+2$ triplets,  the HF solutions representing
the $0^+$ ground states are unambiguously defined by filling the
Kramers-degenerated levels from the bottom of the potential well up
to the Fermi energy.  However, in the odd-$A$  $T=\frac12$ doublets and $A=4n$
$T=1$ triplets,  the HF solutions representing ground states are not
uniquely defined.  The ambiguity is related to the orientation of a
current generated by unpaired nucleon(s) with respect to the nuclear
shape, which gives rise,  in triaxial nuclei, to three different solutions that
have angular momenta oriented along the intermediate ($X$), short
($Y$), and long ($Z$) axes of the nuclear shape~\cite{(Sat12b)}.  No
tilted-axis solutions were found. In  $A=4n$  $|T=1, T_z=\pm 1\rangle
$ states, one should consider,  additionally,  the aligned
$|\nu\otimes \pi\rangle$ (or $|\bar\nu\otimes \bar \pi\rangle$) and
anti-aligned $|\nu\otimes \bar\pi\rangle$ (or $|\bar\nu\otimes
\pi\rangle$) arrangements of the valence neutron ($\nu$) and proton
($\pi$).  Let us mention that both the orientation and alignment
ambiguities are related to the time-odd parts of the functional, which
are very poorly constrained.

To test dependence of MDEs or TDEs on the orientation and alignment
of the HF solutions, for SV$_{\rm T}^{\text{ISB}}$ we performed the complete set of
calculations.  We found that theoretical MDEs in both $T=\frac12$ and
$T=1$ multiplets weakly depend on both the orientation and alignment.
The corresponding uncertainty turned out to be of the order of
10\,keV, and thus in the total uncertainty budget of MDE, it could be
neglected. Accordingly, we performed calculations of MDEs using only
the lowest-energy HF solutions obtained for any given Skyrme
functional.

At variance with the results obtained for MDEs, for TDEs the
influence of the orientations of the HF solutions turned out to be
significant. Hence, in this case  we computed TDEs for each of the
three orientations separately, and then we averaged out the obtained
results. The corresponding standard deviations associated with the
averaging procedure was carried over to the total uncertainty budget.
For SV$_{\rm T}^{\text{ISB}}$, the fitting procedure was performed
for ground-state configurations regardless of their alignment. This
led to anti-aligned configurations in $A = 12$, 20, 24, 36, 40, 44,
48, and 56, and to aligned ones in the remaining $A=4n$ triplets. In
the $A=4n$ triplets, the obtained values of $K$ quantum numbers
(projections of the angular momentum on the symmetry axis) agree with
the experimental values of angular momentum $I$. The only exception
is the $A=12$ triplet, where we obtained $K=2$ for $I=1$.
Furthermore, the procedure was tested for aligned and anti-aligned
configurations separately. These two sets of solutions led to similar
values of \tii coupling constant, and to similar predictions for the
TDEs. Based on this observation, we decided to perform the
calculations for other parametrizations with aligned solutions only,
which are much easier to converge at the HF level.

Unfortunately, for certain values of $A$, we were unable to find
solutions for all orientations. The missing solutions were: for
SV$_{\rm T}^{\text{ISB}}$, $A=12$ and 16; for SkM*$^{\text{ISB}}$,
$A=12$, 16, 20, 24, 28, 32, 40, 44, and 52; for SLy4$^{\text{ISB}}$,
$A=12$, 16, 28, 32, 40, and 52. Furthermore, for SkM*$^{\text{ISB}}$
and SLy4$^{\text{ISB}}$, solutions for $A=30$ and 36 triplets turned
out to be spherically symmetric, which led to unexpectedly large
values of TDE. Therefore, those points were excluded from fitting.

\subsection{The fitting strategy}

Our model has two free coupling constants, \tii and \tiii (see Eqs.\
(\ref{eq:Skyrme_classII})--(\ref{eq:ED_classIII}) of the Letter), which
are adjusted to reproduce all available data on MDEs and TDEs in
isospin doublets and triplets. The most general strategy would be to
perform the 2D fit to the data, that is, to fit both coupling
constants simultaneously. However, the 2D procedure is very costly
computationally, as it requires both the doublets and triplets to be
calculated in the $pn$-mixing formalism. Fortunately, coupling
constants \tii and \tiii are to a large extent independent of one
another. Indeed, test calculations without Coulomb,  performed in
Ref.~\cite{(Bac16a)}, clearly indicate that coupling constant \tii
does not affect MDEs. The same calculations showed that the influence
of coupling constant \tiii on TDEs is also very weak, and comes
mostly from a subtle modification of the HF solution due to the
self-consistency of solutions. This property allows us to replace the
rigorous 2D procedure by two constitutive 1D fits.

\begin{figure}[ht!]
\centerline{\includegraphics[width=0.80\columnwidth]{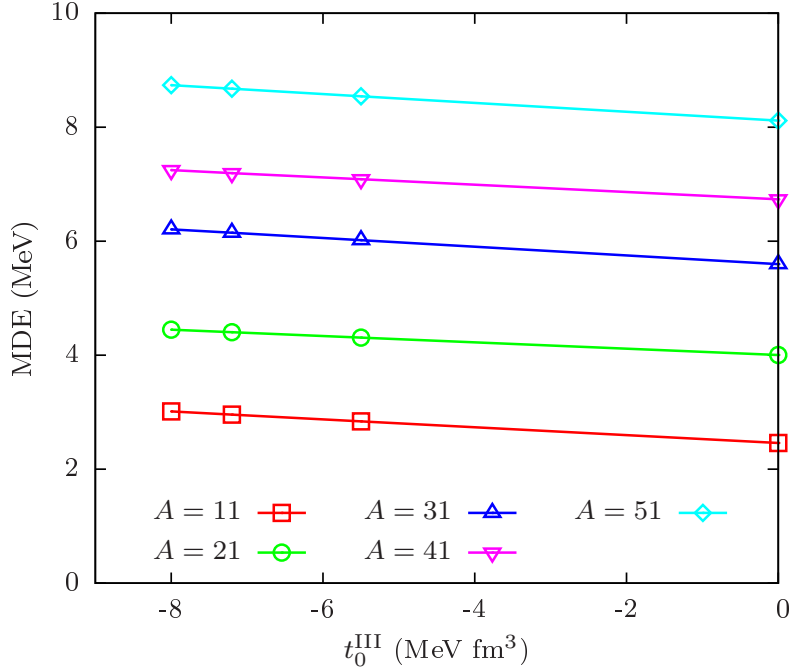}}
\caption{(Color online) MDEs of $T=\frac12$ doublets
calculated for different values of \tiii (\tii = 0) and for SV$_{\rm T}^{\text{ISB}}$.
Calculated points are connected with line
segments that are almost perfectly parallel to one another.}
\label{fig:coefs_mde}
\end{figure}

\begin{figure}[ht!]
\centerline{\includegraphics[width=0.80\columnwidth]{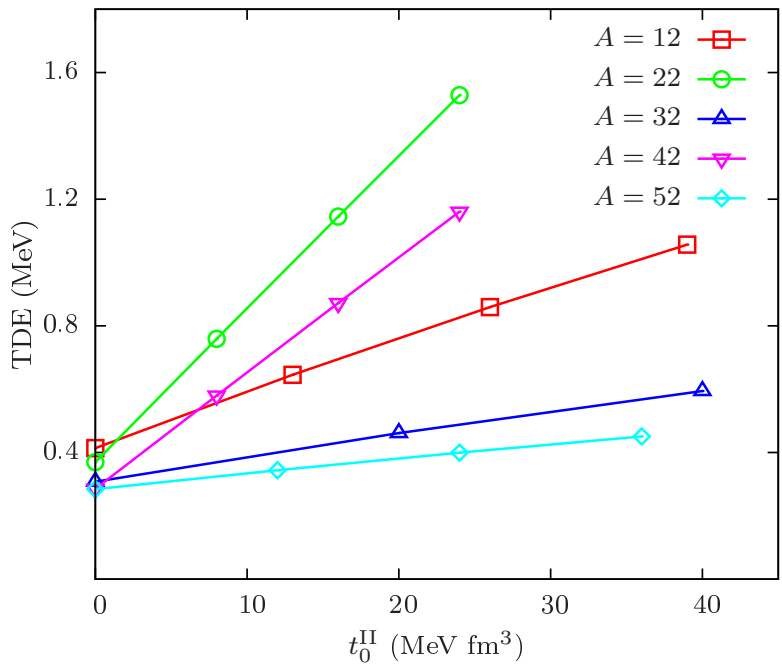}}
\caption{(Color online) TDEs of $T=1$ triplets
calculated for different values of \tii (\tiii=$-$7.4\,MeV fm$^3$)
and for SV$_{\rm T}^{\text{ISB}}$. Calculated points are connected with line
segments.}
\label{fig:coefs_tde}
\end{figure}

To reduce a  possible discrepancy between the full 2D fit and our
simplified fitting procedure, we first performed the adjustment of
\tiii coupling constant to MDEs. In the next step, for the physical
value of \tiii obtained previously,  we performed the adjustment of
\tii  to TDEs.  The adopted strategy  allowed us to  compute the
doublets within the standard $pn$-separable formalism,  thus reducing
the computational effort considerably. Moreover,  the calculations
showed almost perfect linearity of MDEs (TDEs) as functions of \tiii
(\tii). For typical calculations, this fact is illustrated in
Figs.~\ref{fig:coefs_mde} and \ref{fig:coefs_tde}. Such linearity, in
turn, allowed us to compute derivatives of MDEs and TDEs over the
coupling constants by using only two values of the coupling
constants. Actually, to verify the linearity and to double-check the
obtained results, we used three or four values. The errors introduced
by the assumed linearity turned out to be negligible.


\subsection{Penalty function and minimization procedure}

The 1D fitting was based on the $\chi^2$ minimization procedure
described in Ref.~\cite{(Dob14)}. We defined the penalty function
$\chi^2$ in the following way:
\begin{equation}
\chi^2(t_0)=\sum_{i=1}^{N_d}\frac{\left({\rm
DE}_i(t_0)-{\rm DE}_i^{\text{exp}}\right)^2}{\Delta {\rm DE}_{i}^2},
\end{equation}
where $t_0$ represents the actual coupling constants, \tii or \tiii,
used in the 1D fitting, $N_d$ is the number of experimental data
points, DE$_i(t_0)$ represents the displacement energies, MDEs or
TDEs, calculated for a given value of coupling constant $t_0$, and
DE$_i^{\text{exp}}$ are the experimental values of the displacement
energies. The denominator, $\Delta {\rm DE}_{i}^2 = (\Delta {\rm
DE}_{i}^{\text{exp}})^2 + (\Delta {\rm DE}_{i}^{\text{the}})^2 + \Delta {\rm
DE}_{\text{mod}}^2$, is the error comprising three components:
\begin{itemize}
\item $\Delta {\rm DE}_{i}^{\text{exp}}$ is the experimental error evaluated from
errors of the binding and excitation energies as given in
Refs.~\cite{(Wan12a),(ensdf_url)},
\item $\Delta {\rm DE}_{i}^{\text{the}}$ is the theoretical uncertainty related to
the averaging over orientations as described above,
\item $\Delta {\rm DE}_{\text{mod}}$ is the theoretical model uncertainty, which is an
unknown adjustable parameter.
\end{itemize}

The idea of the fit is to minimize $\chi^2$ with the additional
constraint of $\chi^2$ per degree of freedom being equal to one. The
latter condition can be obtained by adjusting the model uncertainty $\Delta {\rm DE}_{\text{mod}}$,
which is responsible for the spread of the
theory-experiment differences. Values of model uncertainties
obtained for three different Skyrme forces used in this work are
collected in Table~\ref{tab:numErr}.
As it turns out, the model uncertainty is the main
source of the total uncertainty, which can be checked by comparing it
with the root-mean-square deviations (RMSD) between theory and experiment, see Table~\ref{tab:numErr}.

\begin{table}[ht!]
\centering
\caption{Values of the model uncertainties, $\Delta {\rm
DE}_{\text{mod}}$, and RMSD of MDEs and TDEs obtained for
three EDFs considered in this work. All values are in keV.}
\label{tab:numErr}
\begin{tabular}{c@{~~~}c@{~~~~}c@{~~~~}c}\hline\rule{0mm}{3.5mm}
 & SV$_{\rm T}^{\text{ISB}}$ & SkM*$^{\text{ISB}}$ & SLy4$^{\text{ISB}}$\\\hline\rule{0mm}{3.5mm}
$\Delta \mathrm{MDE}_{\text{mod}}$ & 190 & 150 & 120\\
RMSD$(\mathrm{MDE})$ & 200 & 150 & 120\\\hline\rule{0mm}{3.5mm}
$\Delta \mathrm{TDE}_{\text{mod}}$ & 63 & 100 & 110\\
RMSD$(\mathrm{TDE})$ & 69 & 100 & 110\\\hline
\end{tabular}
\end{table}

\begin{figure}[hb!]
\centerline{\includegraphics[width=0.8\columnwidth]{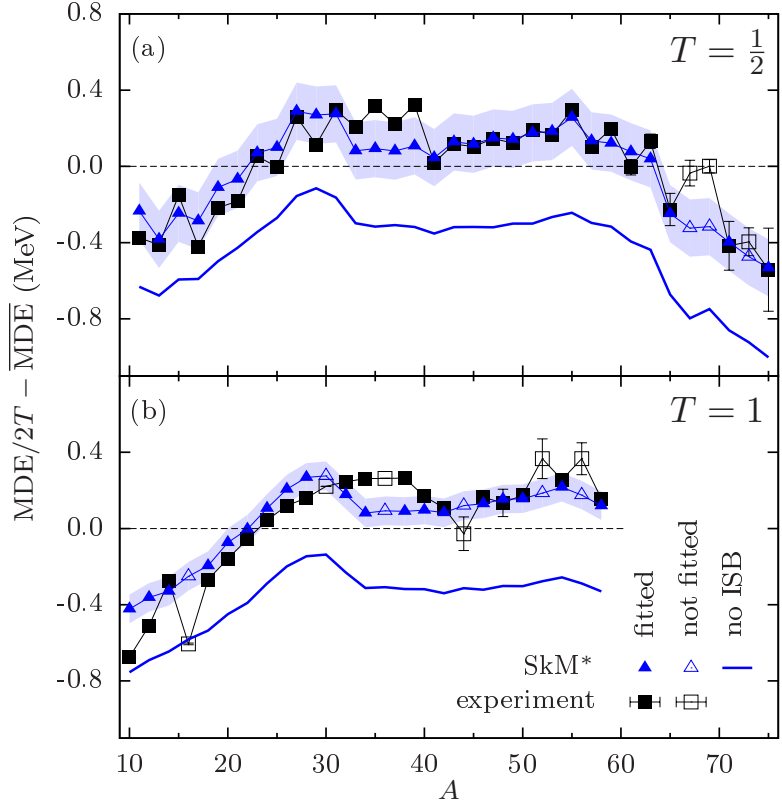}}
\caption{(Color online)
Theoretical and experimental values of MDEs for the $T=1/2$ (a) and $T=1$
(b) mirror nuclei, shown with respect to the average linear trend
defined in Fig.~\protect\ref{fig:MDE_TDE} of the paper. Calculations were
performed using SkM*$^{\text{ISB}}$ functional with the ISB terms added. Shaded bands show theoretical uncertainties. Experimental error bars
are shown only when they are larger than the corresponding symbols. Full (open)
symbols denote data points included in (excluded from) the fitting
procedure.}
\label{fig:MDE_SKM}
\end{figure}

\begin{figure}[hb!]
\centerline{\includegraphics[width=0.8\columnwidth]{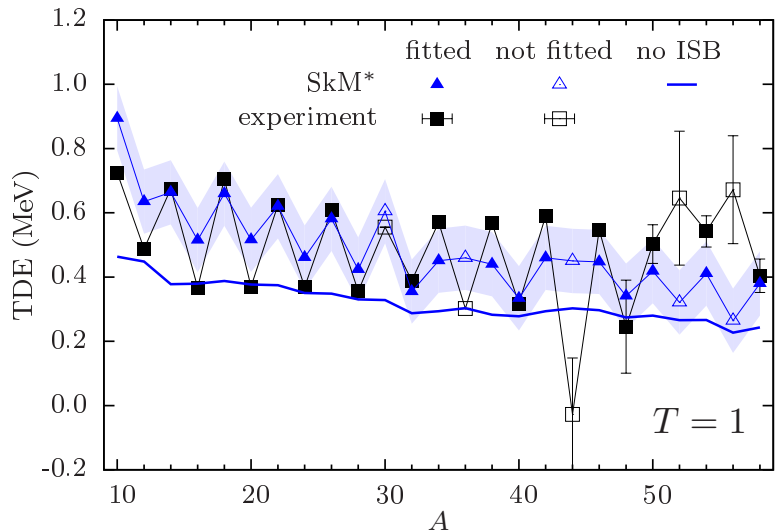}}
\caption{(Color online) Same as in Fig.~\protect\ref{fig:MDE_SKM}
but for the $T=1$ TDEs with no linear trend subtracted.}
\label{fig:TDE_SKM}
\end{figure}

\section{Complementary results for SkM*$^{\text{ISB}}$ and SLy4$^{\text{ISB}}$}

As stated in the Letter, calculations of MDEs and TDEs with the
extended Skyrme forces involving the class-II and III
contact terms were performed for three different Skyrme EDFs SV$_{\rm
T}^{\text{ISB}}$, SkM*$^{\text{ISB}}$, and SLy4$^{\text{ISB}}$. As shown in the Letter, within the model
uncertainties, the resulting coupling constants \tii and \tiii are
fairly independent of the EDF. Below in
Figs.~\ref{fig:MDE_SKM}--\ref{fig:TDE_SLY4}, for the sake of
completeness, we present results obtained for the EDFs SkM*$^{\text{ISB}}$ and
SLy4$^{\text{ISB}}$, whereas the analogous SV$_{\rm T}^{\text{ISB}}$ results were shown in
Figs.~\ref{fig:MDE_SV} and~\ref{fig:TDE_SV} of the Letter. As we can
see, for all three parametrizations, the agreement between
experimental and theoretical values is similar. This nicely confirms
that the method proposed in this work is indeed almost perfectly
independent on the isoscalar charge-independent part of the
functional. In this sense it is a quite robust method
to assess the ISB effects
in $N\simeq Z$ nuclei.

This work was supported in part by the Polish National Science Center
under Contract Nos.\ 2014/15/N/ST2/03454 and
2015/17/N/ST2/04025, by the
Academy of Finland and University of Jyv\"askyl\"a within the FIDIPRO
program,
by Interdisciplinary Computational Science Program in CCS,
University of Tsukuba,
and by ImPACT Program of Council for Science,
Technology and Innovation (Cabinet Office, Government of Japan).
We acknowledge the CI\'S \'Swierk Computing Center, Poland, and the CSC-IT Center for Science Ltd., Finland, for the allocation of
computational resources.

\begin{figure}[ht!]
\centerline{\includegraphics[width=0.8\columnwidth]{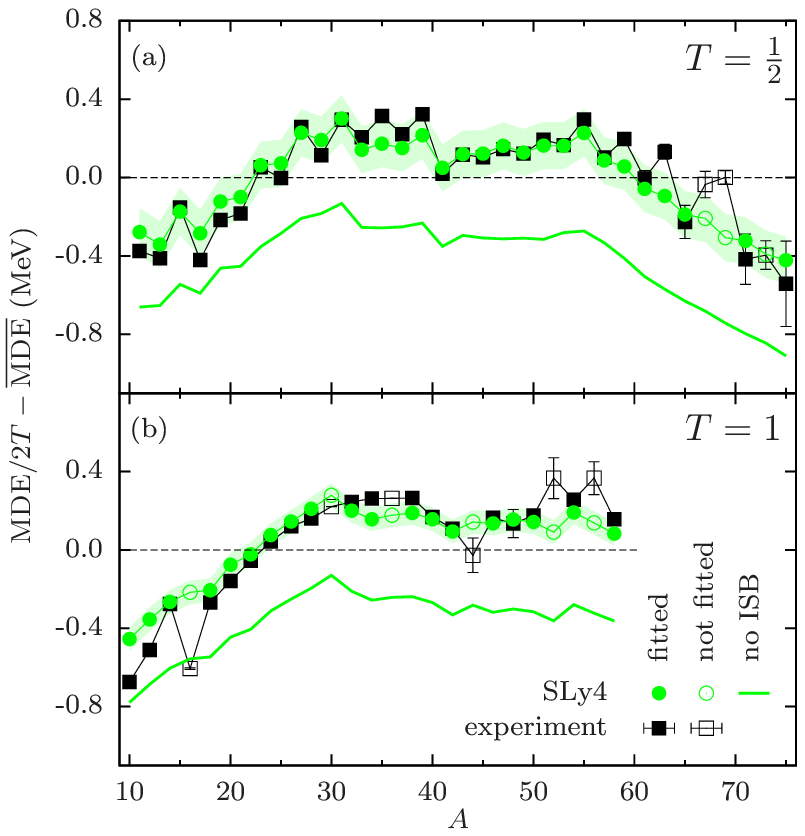}}
\caption{(Color online)
Same as in Fig.~\protect\ref{fig:MDE_SKM} but for the SLy4$^{\text{ISB}}$ EDF}
\label{fig:MDE_SLY4}
\end{figure}

\newpage

\begin{figure}[ht!]
\centerline{\includegraphics[width=0.8\columnwidth]{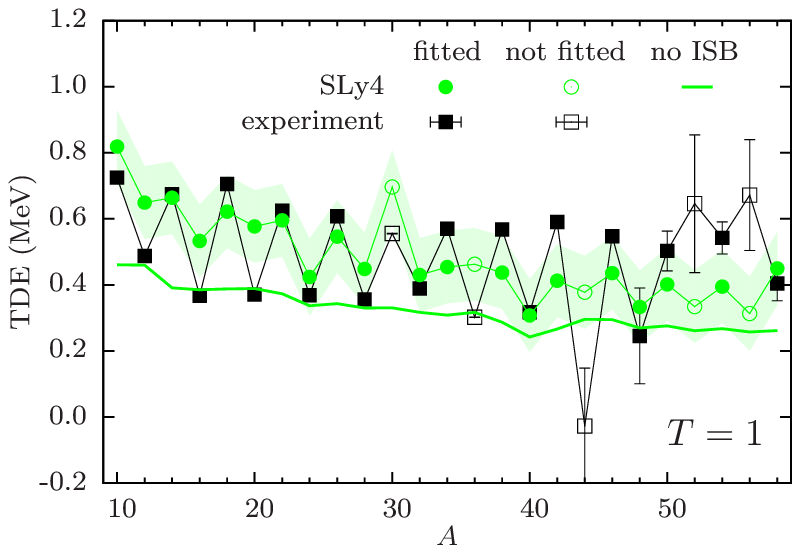}}
\caption{(Color online) Same as in Fig.~\protect\ref{fig:MDE_SLY4}
but for the $T=1$ TDEs with no linear trend subtracted.}
\label{fig:TDE_SLY4}
\end{figure}

\bibliography{TEDMED,jacwit33}
\bibliographystyle{elsarticle-num}